\documentclass[10pt,notitlepage,showpacs,pra,twocolumn]{revtex4-1}
\usepackage{ulem}
\usepackage{amsfonts}
\usepackage{amsmath}
\usepackage{amssymb}
\usepackage{graphicx}
\usepackage{float}
\usepackage[toc,page,header]{appendix}

\begin{document}

\title{Adiabatic and high-fidelity quantum gates with hybrid Rydberg-Rydberg interactions}
\author{Dongmin Yu$^{1}$, Han Wang$^{1}$, Danan Ma$^{1}$, Xing-dong Zhao$^{2}$ and Jing Qian$^{1,\dagger}$ }
\affiliation{$^{1}$State Key Laboratory of Precision Spectroscopy, Department of Physics, School of Physics and Material Science, East China
Normal University, Shanghai 200062, China}
\affiliation{$^{2}$College of Physics and Materials Science, Henan Normal University, Xinxiang 453007, China}

\begin{abstract}
Rydberg blockaded gate is a fundamental ingredient for scalable quantum computation with neutral Rydberg atoms. However the fidelity of such a gate is intrinsically limited by a blockade error coming from a Rydberg level shift that forbids its extensive use. Based on a dark-state adiabatic passage, we develop a novel protocol for realizing a two-atom blockade-error-free quantum gate in a hybrid system with simultaneous van der Waals (vdWsI) and resonant dipole-dipole interactions (DDI). The basic idea relies on converting the roles of two interactions, which is, the DDI serves as one time-dependent tunable pulse and the vdWsI acts as a negligible middle level shift as long as the adiabatic condition is preserved. We adopt an optimized super-Gaussian optical pulse with $k\pi$ ($k\gg 1$) area accompanied by a smooth tuning for the DDI, composing a circular stimulated Raman adiabatic passage, which can robustly ensure a faster operation time $\sim 80ns$ as well as a highly-efficient gate fidelity $\sim0.9996$. This theoretical protocol offers a flexible treatment for hybrid interactions in complex Rydberg systems, enabling on-demand design of new types of effective Rydberg quantum gate devices.
\end{abstract}
\email{jqian1982@gmail.com}
\maketitle
\preprint{}

\section{Introduction}

Quantum logic gate based on neutral atoms has been the core ingredient of quantum computation due to the fact that one qubit can be described by one atom with two hyperfine ground sublevels \cite{Jaksch99,Brennen99,Raimond01,Protsenko02,Sheng18}. Rydberg quantum gate is considered to be a powerful resource with great promises to a wide range of quantum information tasks far beyond the original gate proposals, mainly because of the interatomic strong and tunable interactions, enabling the effective implementations of scalable long-range many-qubit gate \cite{Brion07,Wu10,Su18,Shi18,Levine18} as well as the robust entanglement of individual neutral atoms \cite{Wilk10,Jau16,Zeng17}. 

The remarkable features represented by Rydberg atoms are long lifetime and giant polarizability \cite{Gallagher94}, which results in strong resonant dipole-dipole interaction (DDI) and off-resonant van der Waals interaction (vdWsI), forming the basis for quantum information \cite{Saffman10} and quantum simulation \cite{Weimer10}. As a basic and necessary element for building quantum computers, the two-qubit quantum gate can be realized by the strong long-range interactions enabling conditional excitations of atoms to the Rydberg state, which is so-called blockaded gate as seminally proposed by Jaksch {\it et.al.} \cite{Jaksch00}. This approach can be operated on $\mu$s timescale and does not require a precise control for the Rydberg-Rydberg interaction strength as long as the multiple excitations are fully precluded \cite{Lukin01,Isenhower10,Maller15,Sumanta16,Sun18}, providing more possibilities stretching to scalable multiqubit and ultrafast quantum computation. Unfortunately, a fundamental limitation for such blockaded gate arises from an intrinsic blockade error coming from a Rydberg level shift $V$, which is proportional to $(\Omega/V)^2$ \cite{Saffman05} ($\Omega$ is the effective Rabi frequency) accompanied by a blockade error around $10^{-3}$ \cite{Zhang12}, preventing from its extensive use. 
In the recent years, in order to suppress this blockade error, a rich variety of alternative proposals, such as construction of a novel generalized Rabi frequency by the Rydberg energy shift \cite{Shi17}, using technique of the electromagnetically induced transparency \cite{Mueller09}, or replacing with a long-lifed circular Rydberg state \cite{Xia13} and  a specifically-shaped laser \cite{Goerz14,Su16,Theis16}, have been extensively demonstrated, proving the reduction of gate error beyond the level of $10^{-4}$. Especially, a recent exotic proposal describes an accurate quantum gate by spin echo to suppress the blockade error to the order of $(10^{-3})^2$ by adding two anticlockwise rotations, leading to the intrinsic error of gate beyond the level of $10^{-5}$ which is only limited by the Rydberg state decay \cite{Shi182}.

Alternatively, an adiabatic passage has been developed to be a reliable technique to optical transition driving and phase control between the ground and the interacting Rydberg states \cite{Moller08,Rao14,Wu17}, which can achieve a blockade-error-free quantum gate via F\"{o}rster resonance among nearby different Rydberg levels \cite{Beterov16,Beterov18}. Stimulated Raman adiabatic passage (STIRAP) as one type of adiabatic passages, reveals its robustness to the precise control of coherent population transfer in which the imperfections from experimental perturbation and other preparation defects can be strongly reduced \cite{Bergmann98,Vitanov17}. The remarkable feature in STIRAP comes from its powerful immunity to the population loss as well as the energy shift due to the atom-field detuning from the intermediate state.

Inspired by a recent work \cite{Petrosyan17} where Petrosyan {\it et.al.} presented a study for a two-qubit quantum gate based on adiabatic passage with a sufficient DDI to block the excitation, we propose an expanding protocol for developing a complex but more practical quantum gate where the atoms suffer from simultaneous DDI-type and vdWsI-type Rydberg-Rydberg interactions. The fundamental idea is letting the population transfer adiabatically following a two-atom dark state formed by a pair of counterintuitive circular pulses. Remarkably, we convert the roles of hybrid interactions in circular-STIRAP, {\it i.e.} the DD exchange interaction serves as one of tunable STIRAP pulses and the level shift caused by the vdWsI as a negligible detuning to the intermediate state. In contrast to other blocking schemes the roles of two interactions have been entirely changed here, making it free from any blockade error by interactions. 

In the implementation, we observe that all Rydberg population can be adiabatically transferred and returned back to the input qubit state $|1_c1_t\rangle$ in single pulse circle, without any influence from the strength of vdWsI (middle detuning). The optimized adiabaticity enables the gate error for state $|1_c1_t\rangle$ to be much lower than the optical rotation error purely for the input state $|0_c1_t\rangle$. The intrinsic fidelity error $\mathcal{E}_{11}$ of $|1_c1_t\rangle$ dominated by the Rydberg-state decay as well as the adiabatic error, can be reduced to $\mathcal{E}_{11}\sim4\times 10^{-4}$.

\section{Basic Gate protocol}

\begin{figure}
\includegraphics[width=3.4in,height=3.2in]{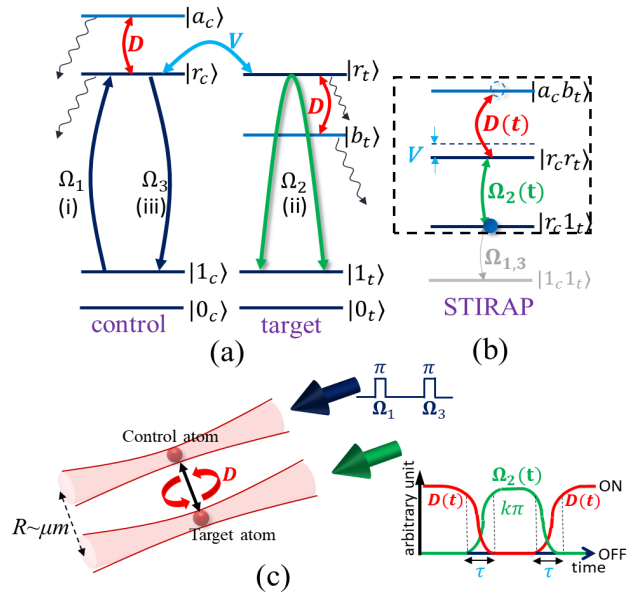}
\caption{(color online). (a) The detailed level descriptions for a two-atom (control and target) quantum gate protocol with the vdWsI labeled by $V$ between identical Rydberg levels and the DDI by $D$ between different Rydberg levels. (b) The initial input two-qubit state $|1_c1_t\rangle$ coupled to $|r_c1_t\rangle$ by $\Omega_{1,3}$($\pi$ pulses) experiences an efficient adiabatic evolution and returns to $|r_c1_t\rangle$ in a circular-STIRAP with the assistance of $\Omega_2(t)$ and $D(t)$ in step ii. The vdWs shift $V$ serves as an intermediate-state detuning to the STIRAP. (c) Experimental setup. Two atoms are localized in optical traps separated by $R\sim\mu$m, driven by the $\pi$-area pulse sequences $\Omega_1$, $\Omega_3$. The pulse $\Omega_2(t)$ combined with $D(t)$ composes a pair of counterintuitive pulses in circular-STIRAP. $\tau$ represents the switching time.}
\label{mod}
\end{figure}

The basic scheme for the two-atom quantum gate consists of identical Rydberg states $|r_{c(t)}\rangle$ and long-lived hyperfine ground states $|0_{c(t)}\rangle$, $|1_{c(t)}\rangle$ as represented in Fig.\ref{mod}(a), allowing the optical transitions of $|1_c\rangle\to|r_c\rangle$ and $|1_t\rangle\to|r_t\rangle$ (others $|0_c\rangle$ and $|0_t\rangle$ are idle). The subscripts {\it c,t} stand for control and target atoms. In addition, two nearby Rydberg levels $|a_c\rangle$, $|b_t\rangle$ are considered. By using a static electric field shifting the energy of state $|a_cb_t\rangle$, the exciting atoms will undergo a resonant DD exchange process (F\"{o}rster resonance) $|r_cr_t\rangle \leftrightarrow|a_cb_t\rangle$ with strength $D(t)$, which were observed and tunable in the time domain by the electric Stark effect in experments \cite{Ryabtsev10,Nipper12,Ravets14,Browaeys16,Comparat10,Carroll04,Vogt06}. Other distant F\"{o}rster effect between nearby hyperfine states is neglected due to the compensation from the F\"{o}rster defect \cite{Petrosyan17}. In addition for identical Rydberg levels $|r_{c}\rangle$ and $|r_{t}\rangle$ it possibly exists a vdWs interaction $|r_cr_t\rangle \leftrightarrow|r_cr_t\rangle$ with strength $V$, shifting the energy of state $|r_cr_t\rangle$. 
An experimental example as described in Fig.\ref{mod}(c) for a real implementation can be given by {\it e.g.} states $|r_c\rangle=|r_t\rangle = |nP_{3/2}\rangle$, $|a_c\rangle=|(n+1)S_{1/2}\rangle$, $|b_t\rangle=|nS_{1/2}\rangle$ of $^{87}$Rb, the effective lifetime of which approximates to hundreds of $\mu s$ for $n\geq 90$, giving rise to the decay rate of Rydberg state around a few kHz \cite{Beterov09}.

The detailed gate procedure is achieved in three steps (i-iii), individually carried out by the laser Rabi frequencies $\Omega_{1\sim3}$ as well as the DDI strength $D(t)$. Two $\pi$-pulse $\Omega_1$ and $\Omega_3$ applied in step i and iii only offer an optical excitation or de-excitation ($|1_c\rangle \leftrightarrow i|r_c\rangle$ or $i|r_c\rangle \leftrightarrow -|1_c\rangle$) for the control atom. $\Omega_2$ acting on the target atom is designed to be a long $k\pi$-pulse and meets $k\gg 2$($k/2$ is an odd integer) for preserving the adiabaticity in STIRAP. This long pulse $\Omega_2$ may cause adiabatic error to our scheme.
In step ii, if the control atom is in the idle state $|0_c\rangle$, applying $\Omega_2$ results in $|1_t\rangle\to-|1_t\rangle$ with a phase accumulation of $k\pi/2$. However, if the control atom is already excited to $|r_c\rangle$ in step i, a proper adjustment for the strong DDI will lead to a smooth adiabatic evolution among states $|r_c1_t\rangle$, $|r_cr_t\rangle$, $|a_cb_t\rangle$, where $D(t)$ and $\Omega_2(t)$ serve as a pair of circular-STIRAP pulses, as represented in Fig.\ref{mod}(b-c). Remarkably, the role of two interactions has been changed. $D(t)$ acts as one pulse in STIRAP, and the level shift caused by the vdWsI between $|r_c\rangle\leftrightarrow|r_t\rangle$ as an intermediate detuning to STIRAP which can be perfectly suppressed by adiabatically following a dark state. Hence we have used the Rydberg energy shifts (two interactions) to realize a perfect circular-STIRAP procedure, rather than to block the excitations, arising the primary advantage of our protocol that is free from any blockade error.

Our gate differs from Petrosyan's work \cite{Petrosyan17} with significant differences. In the former case Petrosyan proposed a two-atom dark state [$\approx|r_c1_t\rangle$] carried out by a weak Gaussian-shaped pulse and a constant DDI strength, requiring the strength of DDI significantly stronger than that of the Gaussian pulse for the transfer to be adiabatic. Hence a large number of population is sustained in the dark state $|r_c1_t\rangle$ without transfer and revival. We adopt a circular-STIRAP performed by a pair of optimal time-dependent pulse sequences, successfully ensuring all population transfer and return back to the input state $|r_c1_t\rangle$ after a single circle, where the relative strength between DDI and the laser field is not important. Such way of counterintuitive optical pulses has been widely used in preparation of ultracold polar molecules, facilitating the manipulation of quantum state and precise measurement \cite{Danzl08,Ospelkaus08,Lang08}.

On the other hand, our circular-STIRAP consists of only one optimized super-Gaussian optical pulse (robustness of a super-Gaussian pulse is provided in Appendix A). The other pulse is played by a smooth tunable DDI between different Rydberg levels. Simultaneously the vdWsI between identical Rydberg levels can be treated as a negligible detuning to the unoccupied middle state $|r_cr_t\rangle$. Remember a traditional blockaded gate relies on a strong interaction strength to block the excitation of multiple Rydberg state although its precise value has no effect \cite{Jaksch00}. To this end, we fully release the requirement for the strengths of both DDI and vdWIs by converting the roles of them. All parameters required in our protocol are feasible by experiment.

\section{Adiabatic passage}

\begin{figure}
\includegraphics[width=3.4in,height=1.8in]{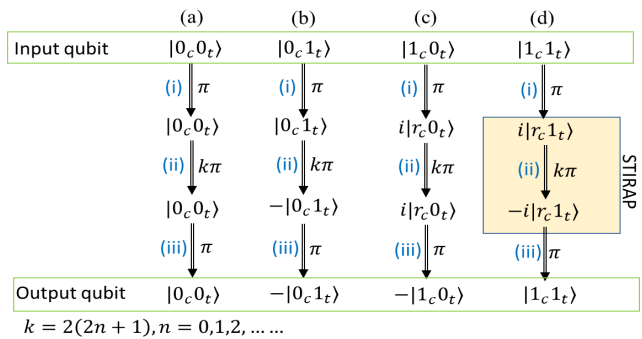}
\caption{(color online). (a-d) The procedure diagram for a two-qubit quantum logic gate protocol by using three pulses $\Omega_1$ ($\pi$ pulse), $\Omega_2$ ($k\pi$ pulse), $\Omega_3$ ($\pi$ pulse). Here the value $k/2=2n+1$ is an odd integer determined by the adiabatic condition in the step ii of (d) which is also called a circular-STIRAP (transfer and revival).}
\label{per}
\end{figure}

For all input qubits $\{|0_c0_t\rangle,|0_c1_t\rangle,|1_c0_t\rangle,|1_c1_t\rangle\}$, the ideal output is $\{|0_c0_t\rangle,-|0_c1_t\rangle,-|1_c0_t\rangle,|1_c1_t\rangle\}$ implemented by optical pulses $\Omega_{1\sim3}$ as schematically shown in Fig.\ref{per}, where the STIRAP works in the step ii of (d) if $|1_c1_t\rangle$ is inputed. Such type of quantum gate can be used as the building blocks of basic quantum circuits, applying for quantum computation and genetic algorithm. 

Supposing all phases of driving Rabi frequencies vanished, so the effective Hamiltonian for a reduced subspace composed of $|r_c1_t\rangle$, $|r_cr_t\rangle$, $|a_cb_t\rangle$ is given by
\begin{eqnarray}
\mathcal{H}_{eff}&=&V\vert r_cr_t\rangle\langle r_cr_t\vert+(\frac{\Omega_2}{2}\vert r_c1_t\rangle\langle r_cr_t\vert  \nonumber \\
         &+&\frac{D}{2}\vert r_cr_t\rangle\langle a_cb_t\vert+H.c.)
\end{eqnarray}
which supports three eigenstates:
\begin{eqnarray}
|d\rangle &=& (D|r_c1_t\rangle -\Omega_2|a_cb_t\rangle)/v \\
|b_+\rangle &=& \frac{\Omega_2|r_c1_t\rangle+(m+V)|r_cr_t\rangle + D|a_cb_t\rangle}{\sqrt{2m(m+V)} } \\
|b_-\rangle &=&  \frac{\Omega_2|r_c1_t\rangle+(m-V)|r_cr_t\rangle + D|a_cb_t\rangle}{\sqrt{2m(m-V)} }
\end{eqnarray}
with $v=\sqrt{D^2+\Omega_2^2}$ and $m = \sqrt{D^2+\Omega_2^2+V^2}$. The corresponding eigenvalues are $E_d = 0$ and $E_{b_{\pm}}=\frac{1}{2}(V\pm m)$. It is clear that $|d\rangle$ is a two-atom dark state that excludes middle state $|r_cr_t\rangle$, resulting in a circular-STIRAP evolution adiabatically following $|r_c1_t\rangle\to|a_cb_t\rangle\to|r_c1_t\rangle$, carried out by the pulse sequence like Fig.\ref{mod}(c). The level spacing between nearby eigenstates can be tuned by the vdWsI, and one finds that increasing or decreasing $V$ leads to a tendency to break the adiabatic condition due to $E_{b_{+}} \to E_d$ or $E_{b_{-}} \to E_d$ if $|V|$ is sufficiently large. At $V=0$ the level spacings are equal to $|E_{b_{\pm}}-E_d| = v$. An explicit expression of adiabaticity can be obtained by exactly solving the sum of all possible population on nearby bright states $|b_{\pm}\rangle$, arising \cite{Pu07}
\begin{equation}
r(t)=\sqrt{\frac{2}{m(m+V)^3}+\frac{2}{m(m-V)^3}}\frac{|\dot{\Omega}_2D-\dot{D}\Omega_2|}{2v}.
\label{rrn}
\end{equation}
Note that if the population of nearby bright states is vanished one can assume all population evolves in the dark state, leading to a perfect adiabaticity under the condition of $r\ll 1$. Additionally, it is remarkable that, from Eq.(\ref{rrn}) the dependence of $r(t)$ on $V$ requires a flexible control of the middle state shift $V$ in a finite range, providing a conditional selection of Rydberg levels. However the real dependence of the maximal value $r^{\max}$ with respect to $V$ is insensitive as verified later.

\begin{figure}
\includegraphics[width=3.0in,height=4.2in]{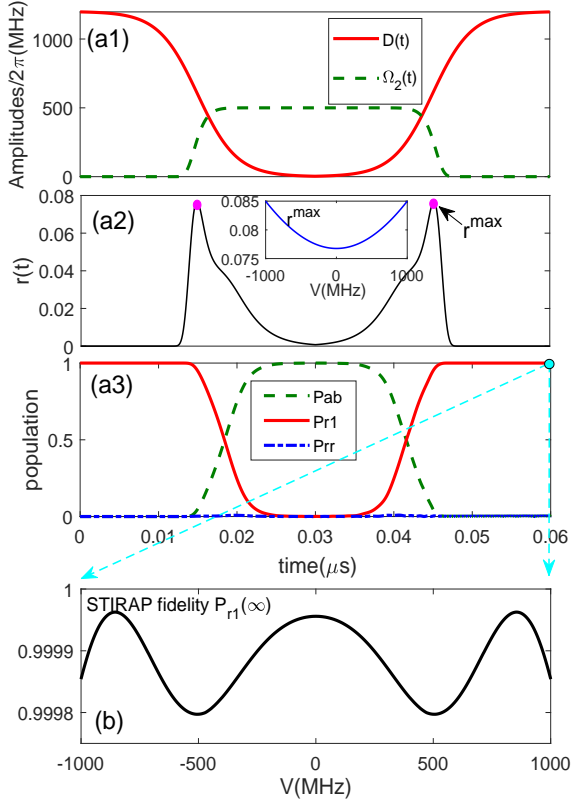}
\caption{(color online). In a circular-STIRAP, (a1-a3) The time dependences of the required pulses $\Omega_2(t)$, $D(t)$, the adiabatic parameter $r(t)$ and the real population transfer $P_{r1}$, $P_{rr}$, $P_{ab}$ (subscripts {\it c,t} are ignored) among these levels. (b) The STIRAP fidelity denoted by $P_{r1}(\infty)$ with respect to the strength of the vdWI for the input qubit state $|r_c1_t\rangle$. }
\label{ap}
\end{figure}

To demonstrate the intrinsic gate error from the adiabatic process in step ii of Fig.\ref{per}(d), we first start from the initial state $|r_c1_t\rangle$ and apply pulses $D(t)$ and $\Omega_2(t)$ with specific shapes that agrees with the real population evolution, as shown in Fig.\ref{mod}(c). An optimization for the specific super-Gaussian pulse shapes $\Omega_2(t)$ in the real implementation is comparably presented in Appendix A with the apply of traditional Gaussian pulses, robustly supporting the improvement under our consideration. Note that such type of shaped pulses are readily accessible by switching ON or OFF the external fields in the time domain. To implement the circular-STIRAP that ensures the returns of all population to $|r_c1_t\rangle$ after all pulses, one only requires a single pair of counterintuitive pulses $\Omega_2(t)$ and $D(t)$, taking forms of 
\begin{equation}
\Omega_2(t) = \Omega_2^{\max}\exp[{-\frac{(t-2t_1)^n}{T^n}}]
\end{equation}
$$ D(t)=\left\{
\begin{aligned}
&\frac{D^{max}}{2}(1-\tanh(\frac{t-t_1}{\tau})),&  t\in[0,2t_1)\\
&\frac{D^{max}}{2}(1+\tanh(\frac{t-3t_1}{\tau})),&  t\in[2t_1,4t_1]
\end{aligned}
\right. 
$$
with the pulse amplitudes $\Omega_2^{\max}$, $D^{max}$, the pulse center $t_1$, the switching speed $\tau$, the pulse width $T$. To induce a perfect adiabatic transfer we apply an optimalized super-Gaussian pulse with smooth shape $n=16$ and sufficient area $A=30\pi$ that persist the adiabatic value $r(t)$ at a low level. Besides, $A=\int_{-\infty}^{\infty}\Omega_2(t)dt$ leads to a quantitative relationship between $T$ and $n$ which is $-T\Gamma(1/n,(\frac{t-t_1}{T})^n)/n=A/\Omega_2^{max}$ with $\Gamma(v,z)$ the incomplete gamma function, ensuring the conversion of population between $|1_t\rangle\leftrightarrow|r_t\rangle$. The expression $D(t)$ adopted is determined by the adiabatic condition following ref.\cite{Ling04} here.

A brief numerical study for the STIRAP quality is carried out under experimental parameters. The resonant DDI strength is $D^{max}\propto C_3/R^3$ with the coefficient $C_3/2\pi=25.82$GHz$\mu$m$^3$ for the principal quantum number $n\geq90$ in Rb atoms \cite{Beterov16}. The two-atom distance $R=2.78\mu$m, leading to $D^{max}/2\pi=1.2$GHz. The time-dependence of $D(t)$ can be carried out by changing the angle $\theta(t)$ between the interatomic axis and the quantization axis \cite{Ravet15}. Assuming $k=30$ ($\Omega_2$ is a 30$\pi$-pulse) results in self-consistent values $\Omega_2^{max}/2\pi=500 $MHz, $t_1=0.015\mu$s, $\tau=0.005\mu$s and $T=0.015503\mu$s. Note that the required $\Omega_2$ is a short ($\sim10ns$) and powerful ($\sim100$MHz) laser pulse, and the resonant DD exchange interaction is sufficient in strong blockaded regime, both accessible by current experimental technique \cite{Huber11,Muller09}.

The spontaneous decay of Rydberg state is $\Gamma/2\pi=0.2$kHz, ensuring $\Omega_2^{max}/\Gamma\sim 10^{6}$. The precise shapes of $\Omega_2(t)$ and $D(t)$ are shown in Fig.\ref{ap}(a1) where two time-dependent pulses occur two switches at time $t=t_1$ and $3t_1$. From Fig.\ref{ap}(a2) we see the adiabatic parameter $r(t)$ sustains a low level $<0.08$ with time, during which the maximal value $r^{\max}$ can reveal a slight change when the vdWI strength $|V|$ is increased to the GHz scale, indicating a good adiabaticity preserved in our scheme. 
The observable quantity is the population of each state $|ij\rangle$ denoted by $P_{ij}(t)$.
The resulting population evolution represents a perfect population transfer between $|r_c1_t\rangle$ and $|a_cb_t\rangle$ in which the intermediate state population $P_{rr}(t)$ obtains a full suppression and other populations $P_{r1}(t)$, $P_{ab}(t)$ adiabatically follow the change of pulses, see Fig.\ref{ap}(a3).

Due to the vanishing of population $P_{rr}$ in the whole transfer process, the STIRAP fidelity characterized by $P_{r1}(\infty)$ has an undisputed and perfect insensitivity [see Fig.\ref{ap}(b)] to the vdWI strength because it only serves as the detuning of $|r_cr_t\rangle$, giving to a powerful evidence for the rationality of neglecting the vdWsI in a real gate operation. Due to the fact that $r^{\max}$ slightly grows with $|V|$, the final STIRAP fidelity $P_{r1}(\infty)$ persists a higher value $\sim$0.9999 with small fluctuations via a wide adjustment for $V$, indicating a flexible operation of Rydberg levels in experiment.

\section{Full Gate performance}

Below, we numerically study the full performance of our two-qubit quantum gate under realistic experimental environment, characterized by the element fidelity $\mathcal{F}_{ij}$ of input qubit $|ij\rangle$ describing the final output probability of target state. We calculate the evolution of two-atom density matrix $\rho$ by solving the master equation $\dot{\rho}=-i[\mathcal{\hat{H}}_0+\mathcal{\hat{H}}_I,\rho]+\mathcal{\hat{L}}_c[\rho]+\mathcal{\hat{L}}_t[\rho]$  with the Hamiltonians of single atom 
\begin{equation}
\mathcal{\hat{H}}_0=(\frac{\Omega_1}{2}|1_c\rangle\langle r_c|+\frac{\Omega_3}{2}|r_c\rangle\langle 1_c|+\frac{\Omega_2}{2}|1_t\rangle\langle r_t|+H.c.)
\end{equation}
and of the hybrid interaction component 
\begin{equation}
\mathcal{\hat{H}}_I=(\frac{D}{2}|r_cr_t\rangle\langle a_cb_t|+H.c.)+V|r_cr_t\rangle\langle r_cr_t|
\end{equation}
with the vdWs shift to the secondary Rydberg state $|a_cb_t\rangle$ ignored by a proper tuning of the F\"{o}rster defect. A calculation of the STIRAP fidelity $P_{r1}(\infty)$ to this shift is provided in Appendix A where we have shown it is readily to obtain a high $P_{r1}(\infty)$ by tuning the value of shift near resonance.

Note that the Hamiltonian is irrelated to $|0_{c,t}\rangle$ due to its uncoupled interaction with all lasers. The effect of spontaneous decay of control and target atoms described by the Lindblad operators $\mathcal{\hat{L}}_c[\rho]$ and $\mathcal{\hat{L}}_t[\rho]$ take forms of 
\begin{equation}
\mathcal{\hat{L}}_k[\rho]=\frac{1}{2}\sum_{j}^{4}[2\mathcal{\hat{L}}_{j,k}\hat{\rho}\mathcal{\hat{L}}_{j,k}^{\dagger}-(\mathcal{\hat{L}}_{j,k}^{\dagger}\mathcal{\hat{L}}_{j,k}\hat{\rho}+\hat{\rho}\mathcal{\hat{L}}_{j,k}^{\dagger}\mathcal{\hat{L}}_{j,k})] 
\end{equation}
with $k\in(c,t)$ and $\mathcal{\hat{L}}_{1,k}=\Gamma|1_k\rangle\langle r_k|$, $\mathcal{\hat{L}}_{2,k}=\Gamma|0_k\rangle\langle r_k|$, $\mathcal{\hat{L}}_{3,k}=\Gamma|1_k\rangle\langle a(b)_k|$, $\mathcal{\hat{L}}_{4,k}=\Gamma|0_k\rangle\langle a(b)_k|$ (if $k=t$, $a$ is replaced by $b$). The two-atom density matrix $\rho$ is a $16\times 16$ matrix with a complete base vector $|\psi\rangle=[$$|00\rangle$, $|01\rangle$, $|0r\rangle$, $|0b\rangle$, $|10\rangle$, $|11\rangle$, $|1r\rangle$, $|1b\rangle$, $|r0\rangle$, $|r1\rangle$, $|rr\rangle$, $|rb\rangle$, $|a0\rangle$, $|a1\rangle$, $|ar\rangle$, $|ab\rangle$$]$ (the subscripts {\it c,t} are ignored). We numerically solve the master equation for the total state vector $|\psi\rangle$ of the system of two four-level atoms. The decay from the Rydberg state will result in the loss leakage of population to the uncoupled states $|0_{c,t}\rangle$, lowering the fidelity of the target state. In calculating the fidelity of each target state with respect to the two-qubit input state $|\phi_i\rangle=\{|00\rangle$, $|01\rangle$, $|10\rangle$, $|11\rangle\}$ we evolute the density matrix $\rho(t)$ with time following the master equation, detecting the four output state at the end of all pulses at $t=80ns$, The observable element fidelity $\mathcal{F}_{ij}$ of input state $|ij\rangle$ is defined as
\begin{equation}
\mathcal{F}_{ij} = \langle\phi_f|\rho(t=t_{end})|\phi_f\rangle
\end{equation}
in which $|\phi_f\rangle = \{|00\rangle$, $|01\rangle$, $|10\rangle$, $|11\rangle\}$ is the four expected target state and $\rho(t=_{end})$ stands for the detected real state after all pulses at $t_{end}=80n$s.

 \begin{figure}
\includegraphics[width=3.4in,height=2.4in]{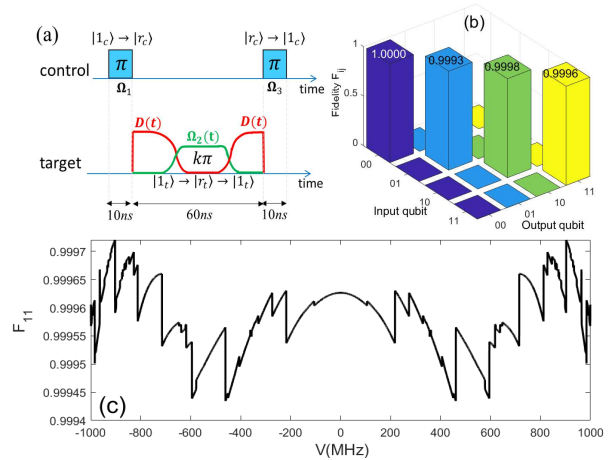}
\caption{(color online). (a) The time diagram of complete pulse sequences $\Omega_{1\sim3}$ and $D(t)$ in a realistic gate operation. $\Omega_1$ and $\Omega_3$ are square pulses with the peak amplitudes 2$\pi\times50$MHz and the durations 10$n$s. The total operation time is as short as $80n$s. (b) The calculated density matrix elements for describing the full gate fidelity $\mathcal{F}_{ij}$ of arbitrary input qubit states $|ij\rangle$. (c) The element fidelity $\mathcal{F}_{11}$ versus the variation of the vdWIs strength.}
\label{gate}
\end{figure}

A specific example for the full gate performance is shown in Fig.\ref{gate} where we present the element fidelity $\mathcal{F}_{ij}$ of input qubit state $|ij\rangle$ utilizing optimal pulse sequences $\Omega_2(t)$, $D(t)$ as well as a pair of square-wave $\pi$ pulses for $\Omega_1(t)$ and $\Omega_3(t)$ separated by a interval of 60$ns$, as plotted in Fig.\ref{gate}(a). The total operation time attains as small as 80$ns$, consisting of two $10ns$-$\pi$ pulses $\Omega_{1,3}(t)$ with amplitudes $2\pi\times50$MHz and a 60$ns$-$k\pi$ circular-STIRAP pulse $D(t), \Omega_2(t)$. Figure \ref{gate}(b) shows a calculated fidelity truth table. For inputs $|10\rangle$ and $|11\rangle$, the element fidelities $\mathcal{F}_{10}$ and $\mathcal{F}_{11}$ sustains as high as 0.9998 and 0.9996, and $\mathcal{F}_{00}=1.0$ because it is irrelevant to all operations and decays. However, a lower value for $\mathcal{F}_{01}$(=0.9993) is observed due to the optical rotation error with the apply of a long $30\pi$-pulse determined by the adiabaticity of STIRAP transfer. Reducing the area of $\Omega_2(t)$ could increase $\mathcal{F}_{01}$ while lowering the value $\mathcal{F}_{11}$ at the same time.

As compared with Fig.\ref{ap}(b), we also re-plot the insensitivity of $\mathcal{F}_{11}$ in the full gate performation with respect to the vdWIs strength, see Fig.\ref{gate}(c). As analogous to $P_{r1}(\infty)$, $\mathcal{F}_{11}$ is still expected to have a perfect insensitivity to $V$ with very small fluctuations $<$0.0003 by adjusting $V$ from -1.0GHz to 1.0GHz, supporting the superiority of our protocol that is free from the vdW blockaded error \cite{Zhang12} because the role of vdWIs has been changed to a negligible middle-state detuning in circular STIRAP. The importance of the optical rotation error will be discussed in section \rm{VB}.

\section{error analysis}

 \subsection{Intrinsic decay and adiabatic errors}

\begin{figure}
\includegraphics[width=3.2in,height=2.9in]{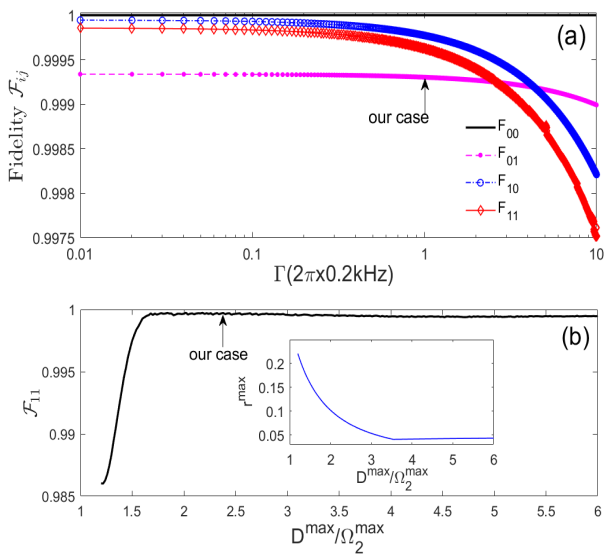}
\caption{(color online). (a) The decay error. A comparison of all elememt fidelities $\mathcal{F}_{ij}$ with respect to the decay rate $\Gamma$. $\mathcal{F}_{00}$, $\mathcal{F}_{01}$, $\mathcal{F}_{10}$, $\mathcal{F}_{11}$ are respectively denoted by black solid, pink with dots, blue with circles, red with diamonds curves. (b) The adiabatic error. In the circular-STIRAP, $\mathcal{F}_{11}$ versus the pulse amplitude ratio $D^{max}/\Omega_2^{max}$. Inset: the maximal adiabatic parameter $r^{\max}$ versus $D^{max}/\Omega_2^{max}$. The arrow points to the special case considered in section \rm{IV}.}
\label{gate2}
\end{figure}

For our gate protocol, the gate error characterized by $\mathcal{E}_{ij}=1-\mathcal{F}_{ij}$ intrinsically comes from two sources. One is the  Rydberg state decay $\Gamma$; the other is the imperfection of adiabaticity, {\it i.e.} the adiabatic error. The former inevitably increases with the value of $\Gamma$. Reducing $\Gamma$ can decrease the decay errors of $\mathcal{E}_{01}$, $\mathcal{E}_{01}$ and $\mathcal{E}_{11}$ except $\mathcal{E}_{00}$, since $|00\rangle$ is isolated from all decays and optical drivings. However for $|01\rangle$, there exists another optical rotation error from a slowly-varying adiabatic pulse $\Omega_2(t)$, only leading to $\mathcal{E}_{01}\to 7\times 10^{-4}$ by a small $\Gamma$. In fact, this rotation error caused by a long pulse can be overcome in a circular-STIRAP scenario when the input qubit state is $|11\rangle$, promising an improving gate fidelity with $\mathcal{F}_{11}>\mathcal{F}_{10}$ when $\Gamma<2\pi\times1.04$kHz. On the other hand, a quantitative comparison with Petrosyan's result is performed by decreasing $\Gamma$ to the level of $\sim2\pi\times0.01k$Hz considered by them, we can obtain comparable gate fidelities $\mathcal{F}_{11}=0.99985$, $\mathcal{F}_{10}=0.99994$ in our scheme.

The latter adiabatic error affecting the gate fidelity $\mathcal{F}_{11}$ only comes from the imperfection of the adiabatic circular-STIRAP transfer, which causes residual population leakage to unwanted intermediate states that does not return back to the target state. As can be numerically studied by changing the relative strength of STIRAP pulse amplitudes $D^{max}$ and $\Omega_2^{max}$, we find that $\mathcal{F}_{11}$ sustains a high value when the coupling strength $D^{max}$ between $|rr\rangle$ and $|ab\rangle$ is large with respect to the pulse amplitude $\Omega_2^{max}$ for keeping the adiabaticity of circular-STIRAP. The fact that $r^{\max}$ decreases rapidly with the increase of $D^{max}/\Omega_2^{max}$ is verified in the inset of Fig. \ref{gate2}(b), which can be understood by the overlap of them. If $\Omega_2^{max}$ is constant, increasing $D^{max}$ could improve the fidelity but still limited by a maximal overlap between $D(t)$ and $\Omega_2(t)$, which makes $\mathcal{F}_{11}$ unvaried when $D^{\max}>1.5\Omega_2^{\max}$. In our calculation, the optimal condition is $D^{\max}/\Omega_2^{\max}=2.4$ as pointed by arrow in Fig. \ref{gate2}(b), greatly different from Petrosyan's work in which $D^{\max}\gg\Omega_2^{\max}$ must be satisfied for the adiabaticity of $|r1\rangle$.

 \subsection{Optical rotation error}

 We propose another way to study the effect of optical rotation error from the long circular-STIRAP pulse $\Omega_2(t)$. In the calculations we have assumed $k=30$(large) for the transfer between $|r1\rangle$ and $|ab\rangle$ to be adiabatic. Such a longer pulse $\Omega_2(t)$ also enables a complete conversion between $|1_t\rangle\leftrightarrow|r_t\rangle$ however bringing extra optical rotation error to $\mathcal{F}_{01}$,  see Fig.\ref{gate2}(a). To illustrate this idea we take a study by lowering the even-integer $k$ from 30 to 2 while other parameters are same as used in section \rm{IV}, in order to see the change of $\mathcal{F}_{01}$ with respect to $k$. With the decrease of pulse area($k\pi$) it is clear that $\mathcal{F}_{01}$ has a linear increase, even arriving $\mathcal{F}_{01}\approx1.0000$ by using a $2\pi$ pulse. It is found that $\mathcal{F}_{10}<\mathcal{F}_{01}$ at $k<10$ because $\mathcal{F}_{10}$ is also affected by a finite time interval ($60\mu$s) between $\Omega_1(t)$ and $\Omega_3(t)$, limited by the Rydberg decay error in this time interval.

\begin{figure}
\includegraphics[width=3.4in,height=2.0in]{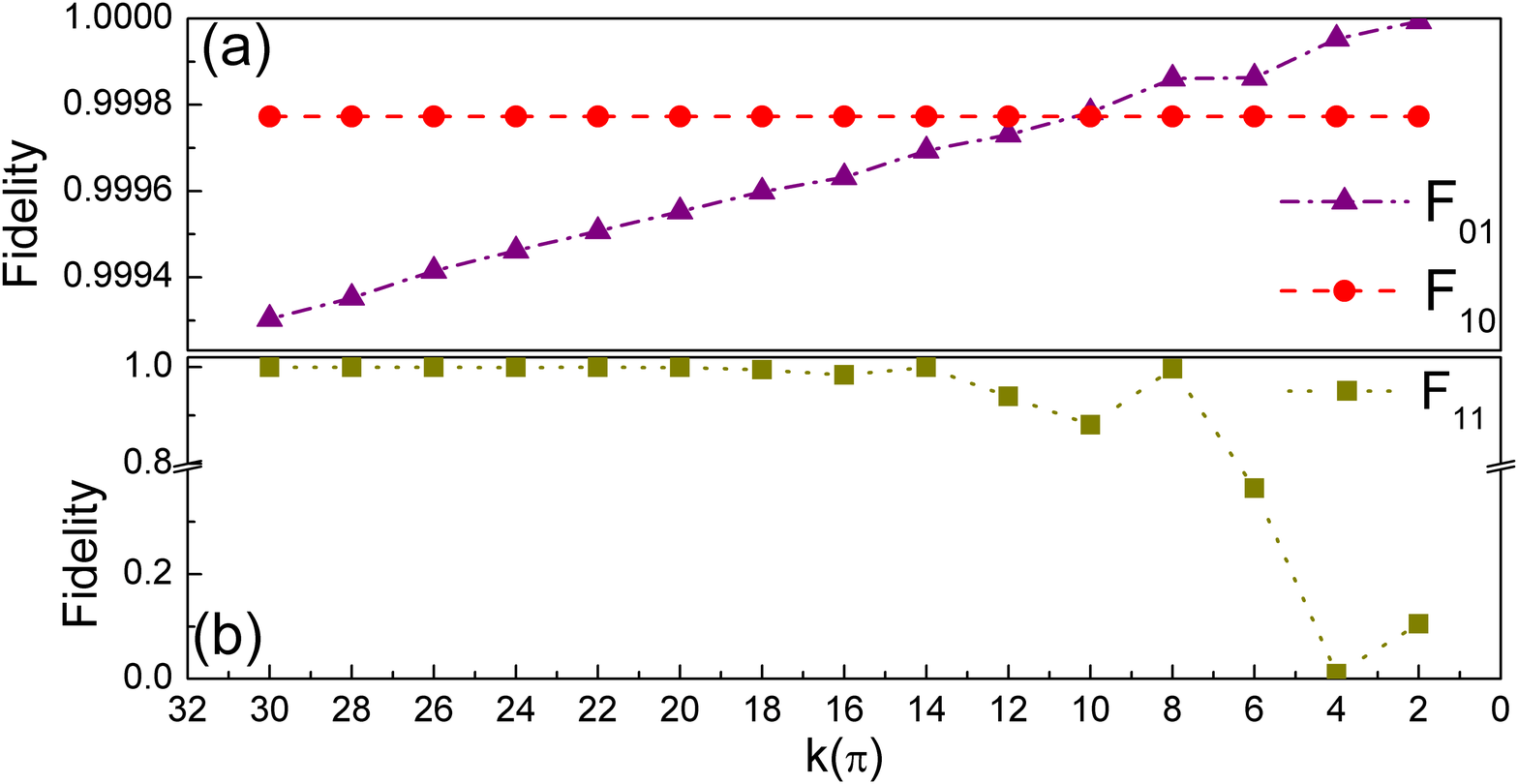}
\caption{(color online). The fidelities (a) $\mathcal{F}_{01}$, $\mathcal{F}_{10}$ and (b) $\mathcal{F}_{11}$ versus the varying the coefficient $k$. $k$ is an even integer and varies in the range of $[2,30]$.  }
\label{errorn}
\end{figure}

For comparison, we plot the fidelity $\mathcal{F}_{11}$ as a function of $k$ in Fig.\ref{errorn}(b). Surely, decreasing $k$ would break the adiabaticity in the circular-STIRAP process, arising a poor fidelity. We see $\mathcal{F}_{11}$ reveals a fast decrease if $k\leq 8$. To this end, it is concluded that in a real gate performance it is important to choose a pulse $\Omega_2(t)$ with a proper area $k\pi$ that can simultaneously sustain the adiabatic transfer for $|11\rangle$, and can reduce the optical rotation error for $|01\rangle$ by optimizing pulse amplitude and pulse duration.

\section{Conclusion}

We explore a novel two-atom quantum gate with hybrid resonant DDI and off-resonant vdWsI based on a perfect circular-STIRAP process. Significantly different from the previous gates, the spirit of our protocol is employing the DDI as a necessary component of STIRAP pulses, and the vdWsI serving as a fully-negligible intermediate-state detuning, so the precise strength of vdWsI is arbitrary in a real gate operation. By using a circular adiabatic excitation of Rydberg levels with time-dependent smooth pulses played by one optical pulse and one DDI pulse, we find the gate fidelity for state $|11\rangle$ that experiences a circular population transfer can even exceed the fidelity of input $|01\rangle$ which is only affected by an optical rotation error from longer $k\pi$ pulse when the Rydberg level is sufficiently long-lifed. The final optimized fidelity can reach $\mathcal{F}_{11}\sim0.9996$ with $\Omega_2^{max}/\Gamma\sim2.5\times10^6$ and the total gate operation time is as smaller as $80n$s under feasible experimental parameters. 
A quantitative comparison with the Petrosyan's work in \cite{Petrosyan17} which requires a sufficiently strong coupling between $|rr\rangle$ and $|ab\rangle$ with respect to the laser pulse amplitude, {\it i.e.} $D^{max}/\Omega_2^{max}\gg1$, for pulses to be adiabatic, we release this requirement by applying an optimal circular-STIRAP with adiabatic and stronger pulses to sustain the adiabaticity, lowering this ratio to $D^{\max}/\Omega_2^{\max}=2.4$, although the absolute values are slightly increased. Furthermore, we note that a circular-STIRAP offers an ideal platform to study the conditional geometric phase shift enabling the yield of controlled phase gate \cite{Wu17}. Analyzing the phase sensitivity in circular-STIRAP with the role of different interactions and shaped pulse fields is importantly a new direction for our future work.

Before ending, we again stress the novelty of our protocol by flexible and realistic operations, in contrast to previous blockade gates accompanied by
the requirement for a strong vdWsI shift. Increasing the vdWI can reduce the intrinsic blockade error but adding extra complexity of gate operation. Currently we propose to convert the roles of hybrid interactions by simply employing an adiabatic passage, relaxing the critical requirements for strong interactions, essentially free from any blockade errors. The flexibility of scheme also relies on an arbitrary determination of vdWsI and DDI between corresponding Rydberg levels, which can be considered as an effective way to implement a two-qubit quantum gate in complex Rydberg systems.

\bigskip

\acknowledgements

This work is supported by the NSFC under Grants No. 11474094, No. 11604086 and No. 11104076,
by the Science and Technology Commission of Shanghai Municipality under Grant No. 18ZR1412800, 
 the ``Fundamental Research Funds for the Central Universities'', the ECNU Academic Innovation Promotion Program for Excellent Doctoral Students YBNLTS2019-023, and the Specialized Research Fund for the Doctoral Program of Higher Education No. 20110076120004.

\appendix

\section{Optimization of pulses in circular-STIRAP}

Here we provide additional information showing the robustness of using a circular-STIRAP with higher-order super-Gaussian pulses, which is essential for revealing advantages of our scheme. We use the subspace composed of states $\{|r_c1_t\rangle,|r_cr_t\rangle,|a_cb_t\rangle\}$ whose effective Hamiltonian can be given by ($\hbar=1$)
\begin{eqnarray}
\mathcal{H}_{eff}&=&V\vert r_cr_t\rangle\langle r_cr_t\vert+\delta\vert a_cb_t\rangle\langle a_cb_t\vert  \nonumber \\
&+&(\frac{\Omega_2}{2}\vert r_c1_t\rangle\langle r_cr_t\vert+\frac{D}{2}\vert r_cr_t\rangle\langle a_cb_t\vert+H.c.)
\end{eqnarray}
For completeness we first assume the vdWs shift $\delta$ with respect to the secondary Rydberg state $|ab\rangle$ is non-zero in Eq.(A1) and study the effect of $\delta$.

In this appendix we numerically calculate the dynamic evolution in which $\Omega_2(t)$ takes a generalized form of
\begin{equation}
\Omega_2(t) = \Omega_2^{\max}e^{-(t-t_2)^{n}/T^{n}}
\label{Ome2}
\end{equation}
whose pulse area satisfies $\int_{-\infty}^{+\infty} \Omega_2(t)dt = k\pi$, arising a relationship for relevant parameters
\begin{equation}
\Omega_2^{\max}T(-\frac{\Gamma(\frac{1}{n},(\frac{t-t_2}{T})^n)}{n})=k\pi
\end{equation}
where $\Omega_2^{\max}/2\pi=500$MHz, $n$ is a positive even integer treating as an order parameter and $T$ is $n$-dependent pulse width, $k=30$(constant), $\Gamma(v,z)$ is incomplete gamma function solved numerically. If $n=2$ it is a Gaussian-type pulse; while if $n\geq4$, $\Omega_2(t)$ is a super-Gaussian function enabling the simulation of a more realistic operation by external fields manipulation.

\begin{figure}
\includegraphics[width=3.4in,height=3.8in]{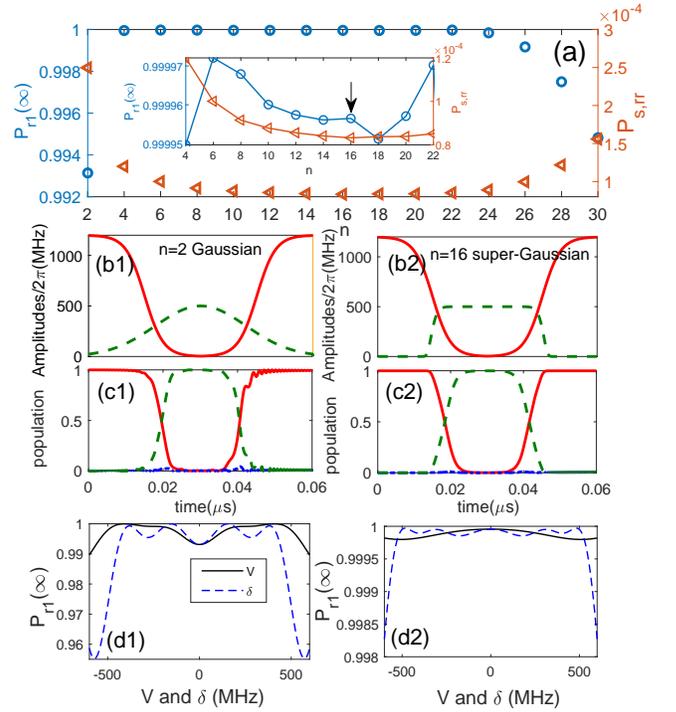}
\caption{(color online). A comparisom of population dynamics between Gaussian-type and super-Gaussian-type pulses in the circular-STIRAP process. (a) The STIRAP fidelity $P_{r1}(\infty)$ and the accumulated population $P_{s,rr}=\int_{-\infty}^{+\infty} P_{rr}(t)dt$ of $|rr\rangle$ are shown, versus the change of order parameter $n$ (an even integer). Inset of (a) is an enlarged frame covering a smaller range of $n\in[4,22]$. When $\Omega_2(t)$ is a Gaussian pulse, (b1) The time-dependence of the applied pulses $\Omega_2(t)$(green-dashed) and $D(t)$(red solid), (c1) The real time-dependent population dynamics $P_{1r}(t)$(red solid), $P_{ab}(t)$(green dashed), $P_{rr}(t)$(blue dotted) and (d1) The dependence of STIRAP fidelity $P_{r1}(\infty)$ on $V$ (black solid) and $\delta$ (blue dashed) are respectively presented. Similar results are shown in (b2-d2) except for a super-Gaussian pulse with $n=16$.}
\label{ad2}
\end{figure}

To order to obtain an optimal value $n$ that can improve the performance of circular-STIRAP, we give the optical Bloch equations in the subspace with vectors $|i\rangle=\{|r1\rangle,|rr\rangle,|ab\rangle\}$ and $i\in(1,2,3)$, given by
\begin{eqnarray}
&\dot{\rho}_{11}& =\gamma(\rho_{22}+\rho_{33})+\frac{i\Omega_2}{2}(\rho_{12}-\rho_{12}^*) \nonumber  \\
&\dot{\rho}_{22}& = -\gamma\rho_{22}+iD(\rho_{23}-\rho_{23}^*)-\frac{i\Omega_2}{2}(\rho_{12}-\rho_{12}^*) \nonumber \\
&\dot{\rho}_{33}& =-\gamma\rho_{33} -iD(\rho_{23}-\rho_{23}^*)   \nonumber\\
&\dot{\rho}_{12}& = (-\frac{\gamma}{2}+iV)\rho_{12}+iD\rho_{13}+\frac{i\Omega_2}{2}(\rho_{11}-\rho_{22}) \\
&\dot{\rho}_{13}& =(-\frac{\gamma}{2}+i\delta)\rho_{13}+iD\rho_{12}-\frac{i\Omega_2}{2}\rho_{23} \nonumber \\
&\dot{\rho}_{23}& =-(\gamma+i(V-\delta))\rho_{23}-iD(\rho_{33}-\rho_{22})-\frac{i\Omega_2}{2}\rho_{13}  \nonumber
\end{eqnarray}
where diagonal elements $\rho_{11}$, $\rho_{22}$, $\rho_{33}$ denoted by $P_{r1}$, $P_{rr}$, $P_{ab}$ means the population, and non-diagonal elements $\rho_{12}$, $\rho_{13}$, $\rho_{23}$ and their conjugated components stand for coherence of transitions.

By directly solving Eqs.(A4) we can extract the information about the dependence of the final STIRAP fidelity denoted by the population
$P_{r1}(\infty)$ (blue circles) on the order parameter $n$ of the applied pulses as shown in Fig.\ref{ad2}(a1). 
In the same picture, we calculate the sum of possible accumulated population in the middle state $|rr\rangle$, labeled by $P_{s,rr}$ (red triangles), revealing that $P_{s,rr}$ can be as low as $\sim10^{-4}$ for all $n$ values. Figure \ref{ad2}(a1) promotes us to search for a determined $n$ that both $P_{r1}(\infty)$ and $P_{s,rr}$ can be optimized. Compared with the case of $n=2$(Gaussian pulse) that suffers from a poor output because of a slightly big accumulation in middle state, we find that increasing the order parameter $n$ in a finite range can significantly enhance $P_{r1}(\infty)$ from 0.993($n=2$) to 0.99997($n=6$). On the other hand, $P_{s,rr}$ has a slow decrease with the increase of $n$. To this end, we choose $n=16$(optimal) in the full gate performation that can preserve simultaneously a high STIRAP fidelity $P_{r1}(\infty)\approx0.99996$ as well as a small middle-state occupancy $P_{s,rr}\approx8.3\times10^{-5}$. 

Figure \ref{ad2}(b1-c1) and (b2-c2) present the respective pulse profiles and the population dynamics for $n=2$ and $n=16$. Correspondingly, (d1-d2) show the dependence of $P_{r1}(\infty)$ with respect to the middle vdWs shift $V$(black solid) and the upper-level shift $\delta$(blue-dashed). It is observed that the insensitivity to $V$ is relatively robust irrespective of $n$, although its fluctuations[$\sim0.01$] is slightly bigger for $n=2$. However, for the upper vdWs shift $\delta$ of $|ab\rangle$ this insensitivity becomes poor, especially note that the apply of a sufficient $|\delta|$ will cause that $P_{r1}(\infty)$ falls down quickly due to the breakup of two-photon resonance. 
When $|\delta|\gg0$ the dark-state adiabatic evolution following $|d\rangle$ can no-longer be persisted, giving rise a big fall of $P_{r1}(\infty)$ irrespective of the value $n$ there. Proper tuning of the F\"{o}rster defect can compensate this shift $\delta$ so we ignore the effect of $\delta$ in the text, see Eq.(1). Surely our results have verified that the insensitivity of $P_{r1}(\infty)$ to the modulus of $\delta$ in a near-resonant regime is still well kept, providing more flexible selections of nearby Rydberg levels $|a_c\rangle$, $|b_t\rangle$ in the gate operation.

\end{document}